\documentclass[twoside]{pro12}
\usepackage[ansinew]{inputenc}
\usepackage{amsmath}
\usepackage{graphicx}
\usepackage{lineno}

\head{ L. Lamy}{SKR southern and northern rotational modulations }

\begin{document}

\title{VARIABILITY OF SOUTHERN AND NORTHERN SKR PERIODICITIES}
\author{L. Lamy\adress{\textsl{Laboratoire d'Etudes Spatiales et d'Instrumentation en Astrophysique, Observatoire de Paris, CNRS, Meudon, France}}$\,$}

\maketitle

\begin{abstract}
Among the persistent questions raised by the existence of a rotational modulation of the Saturn Kilometric Radiation (SKR), the origin of the variability of the 10.8~hours SKR period at a 1\% level over weeks to years remains intriguing. While its short-term fluctuations (20-30 days) have been related to the variations of the solar wind speed, its long-term fluctuations (months to years) were proposed to be triggered by Enceladus mass-loading and/or seasonal variations. This situation has become even more complicated since the recent identification of two separated periods at 10.8h and 10.6h, each varying with time, corresponding to SKR sources located in the southern (S) and the northern (N) hemispheres, respectively. Here, six years of Cassini continuous radio measurements are investigated, from 2004 (pre-equinox) to the end of 2010 (post-equinox). From S and N SKR, radio periods and phase systems are derived separately for each hemisphere and fluctuations of radio periods are investigated at time scales of years to a few months. Then, the S phase is used to demonstrate that the S SKR rotational modulation is consistent with an intrinsically rotating phenomenon, in contrast with the early Voyager picture.
\end{abstract}

\section{Introduction}

The Saturn Kilometric Radiation (SKR) is an intense non-thermal radio emission produced by auroral electrons moving along magnetic field lines, predominantly on the dawn sector [Kaiser et al., 1980]. Its regular pulsation, whose origin still remains unexplained, was originally interpreted as a clock-like rotational modulation triggered by the planetary magnetic field, and thus directly relating to the planetary interior. The Voyager determination of the SKR period (10h39min24$\pm$7s or 10.657$\pm$0.002h) [Desch and Kaiser,1981] was adopted as the official rotation rate of the planet by the International Astronomical Union [Seidelmann et al., 2001], and used to define the Saturn Longitude System (or SLS, hereafter named SLS 1). In this system, the SKR occurrence or intensity is organized by sub-solar longitudes and peaks at a fixed arbitrary value (historically set at $100^\circ$).

Ulysses observations later showed that the observed SKR period ($\sim$10.8h) is not constant but varies with time by 1\% over years [Galopeau and Lecacheux, 2000], which was confirmed by Cassini measurements [Gurnett et al., 2005]. Further studies identified short-term 20-30 days oscillations of the southern SKR period [Zarka et al., 2007] correlated to the fluctuations of the solar wind speed (while the SKR intensity is correlated to the solar wind dynamic pressure) supporting previous modeling work [Cecconi and Zarka, 2005], together with long-term yearly oscillations also found in magnetospheric plasma and magnetic field data and attributed to Enceladus mass-loading [Gurnett et al., 2007]. New longitude systems (SLS 2,3), based on polynomial fits of the SKR long-term phase variations successively replaced SLS 1 [Kurth et al., 2007, 2008]. 

The identification of a second SKR period around 10.6h [Kurth et al., 2008] has strong implications on the validity of the longitude systems described above. Indeed, radio periods at $\sim$10.8h and $\sim$10.6h were respectively attributed to SKR emissions emanating from the southern (S) and the northern (N) hemispheres [Gurnett et al., 2009]. It is interesting to note here that Voyager essentially observed northern SKR, while its orbital motion out of the ecliptic enabled Ulysses to observe both hemispheres close to Saturn's equinox. As both Cassini-derived SKR periods vary with time and come close together about Saturn's equinox (Aug. 2009), these authors proposed that seasonal variations of the solar illumination may induce a different magnetosphere-ionosphere coupling in each hemisphere, and trigger different S,N radio periods varying in opposite ways, with an expected crossing about equinox, when the sun illuminates both hemispheres equally.


Section \ref{technique} describes the employed radio dataset and the periodogram technique used to derive separate S and N SKR periodicities as a function of time. Section \ref{variability} investigates fluctuations of these periods at time scales of years to a few months, and implications on magnetospheric dynamics. Long-term phase systems are then built for each hemisphere. Finally, section \ref{modulation} investigates the organization of S SKR sources as a function of S phase.

\section{Radio observations and harmonic analysis}
\label{technique}

\subsection{Northern and southern SKR emissions}

Taking advantage of quasi-continuous SKR observations acquired by the Cassini Radio and Plasma Wave Science (RPWS) experiment since 2004, radio data were processed as detailed by Lamy et al. [2008a] to obtain regular dynamic spectra of circular polarization degree V and radiated power P (in W.sr$^{-1}$) from 3.5 to 1500~kHz between 1 January 2004 (DOY 1) and 19 October 2010 (DOY 2484) with a 3~min time resolution. A time series of the total radiated power P$_\mathit{SKR}$(t) was  then obtained by integrating P over the spectral range 40 to 500~kHz, that corresponds to well-defined SKR bursts and avoids narrowband low frequency emissions below 40~kHz, that display periods comparable to SKR ones but with a significant phase shift [Ye et al., 2010].

Southern (S) and northern (N) radio emissions were separated on the basis of (i) the SKR property to be primarily emitted on the extraordinary mode [Kaiser et al., 1984; Lamy et al., 2010 and references therein], that displays left-handed (LH, V~$\ge$~0) and right-handed (RH, V~$\le$~0) circular polarization for S and N sources, and (ii) its visibility domain, that mainly illuminates its hemisphere of origin and extends down to 20$^\circ$ latitude in the other hemisphere (see Fig. 11 of Lamy et al., [2008a]). S and N SKR emissions were thus identified by LH and RH emissions respectively observed from latitudes $\lambda_\mathit{sc}\le20^\circ$ and $\lambda_\mathit{sc}\ge-20^\circ$, then integrated from 40 to 500~kHz to build time series P$_\mathit{SKR,S}$(t) and P$_\mathit{SKR,N}$(t). This technique permits one to consider all near-equatorial observations where S and N SKR are observed together (most of Cassini measurements), generally excluded by separating S,N SKR from a single geometrical selection on the spacecraft latitude.

\begin{figure}[ht]
\centering
\includegraphics[width=\textwidth,angle=0]{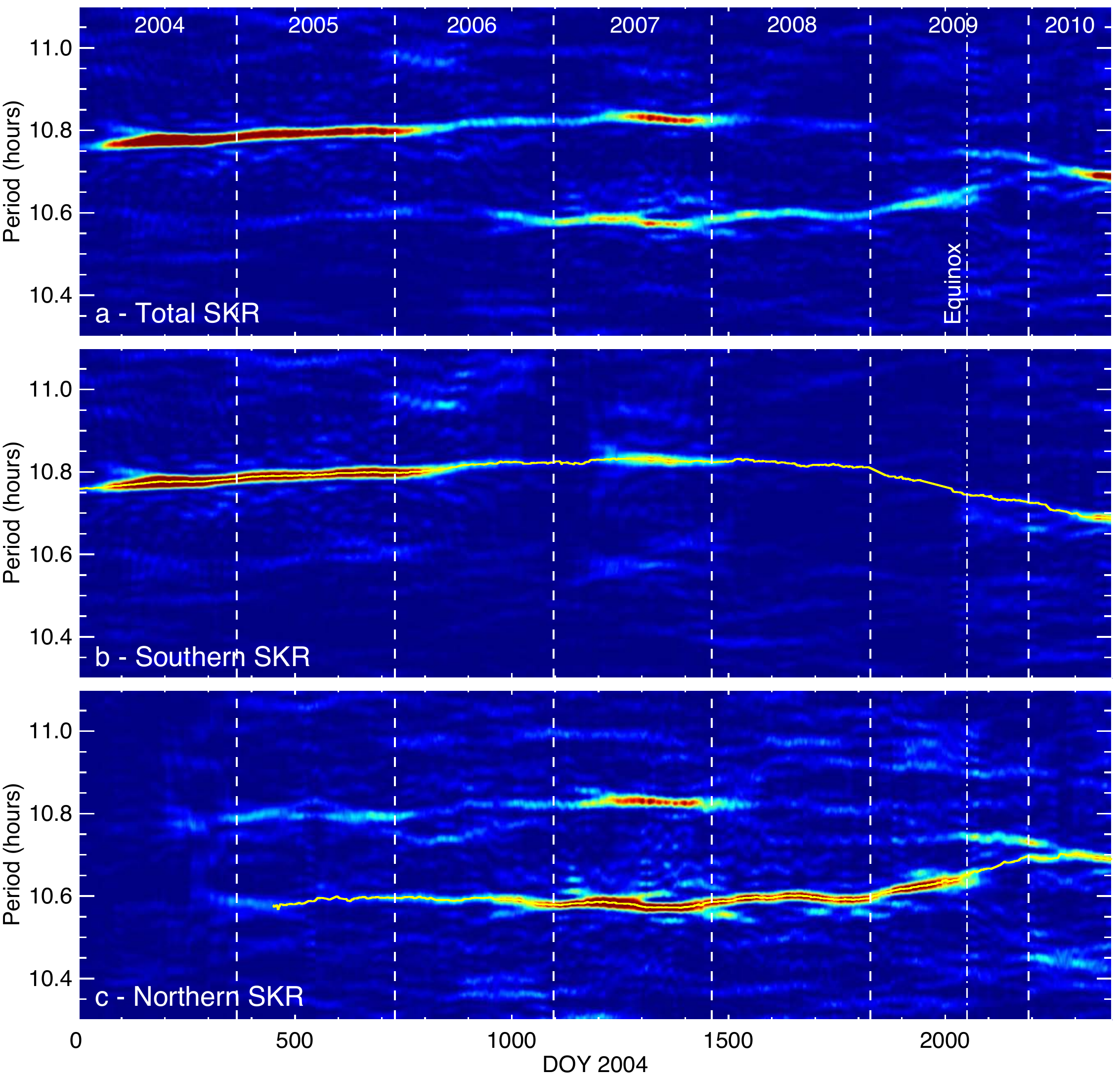} 
\caption{Lomb-Scargle 2D periodograms of (a) $\log$~P$_\mathit{SKR}$(t), (b) $\log$~P$_\mathit{SKR,S}$(t) and (c) $\log$~P$_\mathit{SKR,N}$(t) as a function of period and time. Individual power spectra (columns, in arbitrary units) were computed everay day over a 200-days long window. Solid yellow lines show S and N periods, as derived from panels (b) and (c).}
\centering
\label{fig1}
\end{figure}

\subsection{Lomb-Scargle normalized periodogram}

The Fast Fourier Transform (FFT) is the technique employed most often to perform the spectral analysis of regular time series. However, it cannot deal with unevenly spaced data and, when applied to discontinuous series re-interpolated on a regular basis, it yields noisy power spectra with a significant spectral leakage. The Lomb-Scargle (LS) normalized periodogram is a powerful alternate technique directly applicable to irregularly sampled data that possesses useful properties [Lomb, 1976; Scargle, 1982; Horne and Baliunas, 1986]. The periodogram analysis is equivalent to least-squares fitting of weighted sine curves to the data. Its normalization by the total variance leads to less noisy power spectra and, while the FFT spectral resolution is fixed by the time step of the original time series, the periodogram can be oversampled in order to improve the determination of significant spectral peaks. Due to non strictly continuous RPWS measurements together with the applied data selection, regular SKR power series described above intrinsically contain unevenly spaced non-null signals, for the spectral analysis of which the LS technique is particularly adapted. 


\section{Variability of SKR periodicities}
\label{variability}

\subsection{Long-term variations}
\label{longterm}

Figure \ref{fig1} displays 2D periodograms of the LS power spectra of $\log$~P$_\mathit{SKR}$(t), $\log$~P$_\mathit{SKR,S}$(t) and $\log$~P$_\mathit{SKR,N}$(t) as a function of period and time, computed with a 200-days long sliding window and a time resolution of 1 day. Figure \ref{fig1}a reveals the two main SKR periods around 10.8h and 10.6h, each slowly varying with time, corresponding to southern and northern SKR sources. This is illustrated in Figures \ref{fig1}b and \ref{fig1}c, where S and N SKR mainly pulse at periods about 10.8h and 10.6h, identified by solid yellow lines hereafter labelled $T_\mathit{S,N}$(t). In Figure \ref{fig1}a, the intensity of the S peak is dominant over 2004-2007 and then decreases relative to the intensity of the N peak when approaching the equinox. This reflects the dominant observed signal which corresponds, for comparable observing conditions of both hemispheres (for instance at the equator), to the intrinsic SKR power radiated by each of them, itself observed to change with seasons: while N SKR was more intense during northern summer at the Voyager epoch [Kaiser et al., 1984], S SKR was predominant during the first years of southern summer observed by Cassini [Lamy et al., 2008a]. 

\begin{figure}[ht]
\centering
\includegraphics[width=0.9\textwidth,angle=0]{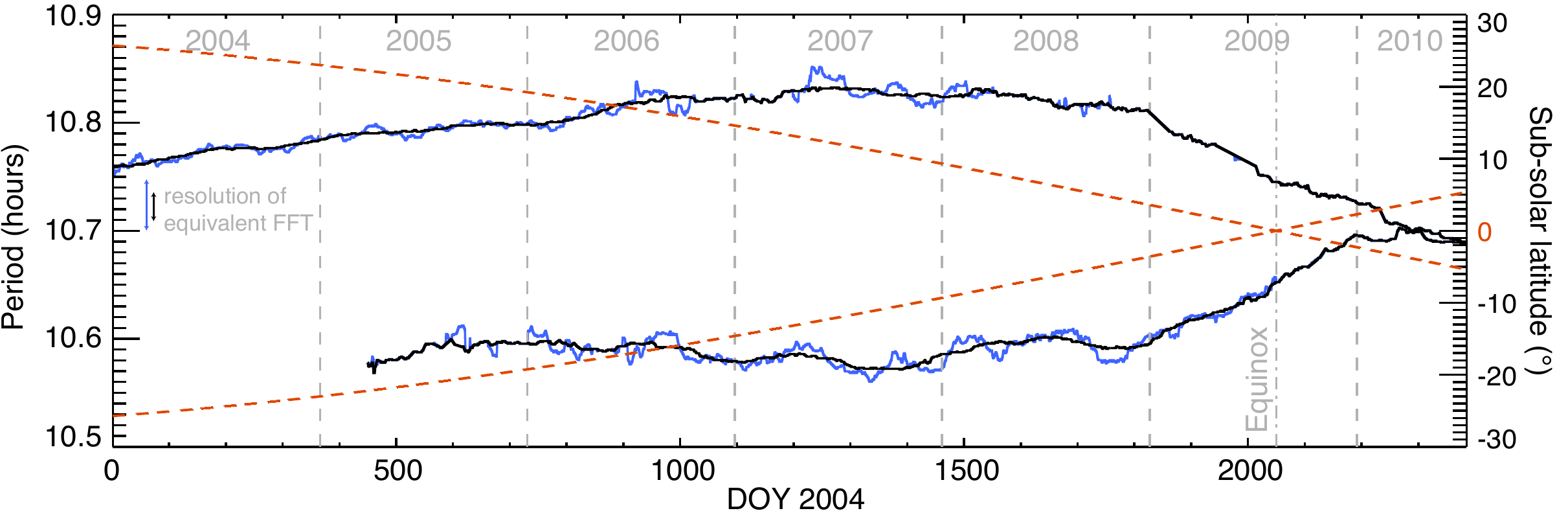} 
\caption{S and N SKR periods derived from a Lomb-Scargle analysis of SKR power using 200-days long (black, Figures \ref{fig1}b,c) and 100-days long (blue) sliding windows. The equivalent resolution of FFT is indicated by double arrows on the left side of the Figure. Both periods cross at day 2010-097 (DOY 2289). Orange dashed curves indicate the sub-solar latitude $\lambda_\mathit{sun}$ and $-\lambda_\mathit{sun}$ (right axis) that cross at the equinox of 11 Aug. 2009 (2009-223 or DOY 2050).}
\centering
\label{fig2}
\end{figure}

$T_\mathit{S}$(t) and $T_\mathit{N}$(t) are plotted together in Figure \ref{fig2} (black lines). They display clear opposite trends, with a correlation coefficient reaching $c=-0.95$, maximal for a lag of 0 days. Both periods cross about 7 April 2010 (DOY 2050), namely approximately 8 months after the equinox. The S period is maximal around 16 June 2007 (DOY 1263) and the N one after 2006 is minimal around 20 October 2007 (DOY 1389). Out of this interval, both periods significantly shift from their extrema.

This anti-symmetrical behaviour supports a seasonal origin. However, the delay of 8 months between the equinox and the actual reversal of both periods indicates a non-linear response of the magnetosphere. Focussing on seasonal variations of the solar illumination, Gurnett et al. [2009] proposed that the temporal variation of Saturn's inclination controls the atmospheric Pedersen conductivity in each hemisphere, which in turn triggers the strength of associated auroral field-aligned current systems, and thus the intensity of the torque (opposite to corotation) at the footprint of the magnetic field lines where these currents close. As the Pedersen conductivity varies quasi-instantaneously with the solar illumination, this model needs to be augmented with a significant mechanical slippage responsible for the observed lag. Also, while they are close to each other (within 100 days), the extrema of both periods are shifted with respect to the ones of the solar illumination (dashed lines in Figure \ref{fig2}) by approximately 2 years, suggesting that additional effects shall affect the variation of radio periods. 


Interestingly, N SKR displays a secondary peak at the S period (obvious in 2005 and 2007). Although it cannot be formally excluded, it is unlikely that such an intense secondary peak results from a non-ideal data selection with a possible residual contribution of S SKR sources to P$_\mathit{SKR,N}$(t). Moreover, this secondary peak remains when separating S and N emissions from observing latitudes above thresholds of $10^\circ$ or 20$^\circ$, and is supported by equatorial observations (see example in 2007 displayed by Figure \ref{figsup}), where part of the N emission pulses together with the S one. This result is thus likely to reflect a real physical dual modulation. This is not surprising since SKR sources have been identified on closed field lines [Lamy et al., 2010; Bunce et al., 2010], and that auroral electrons accelerated in one hemisphere could ultimately reach the other one, as proposed for interpreting Io's multiple footprints [Bonfond et al., 2008]. Moreover, both periodicities have been detected in northern electron observations [Carbary et al., 2009], as well as in equatorial magnetic field oscillations [Provan et al., 2011].
 

In a parallel paper, Gurnett et al. [2010] specifically investigate the crossing of SKR periods. They show that periodicities observed by Ulysses (determined by FFT) are consistent with the observation of S,N hemispheres, and crossed about 9 months after the equinox of 1995. With a different technique of period determination, they also identified a crossing of both SKR periods from Cassini data approximately 7 months after the equinox of 2009. Although they are derived with a 240-days sliding window excluding near-equatorial observations, these periods are roughly consistent with the ones derived above. However, the 8 months delay measured from the present LS analysis is in better agreement with the one derived from Ulysses data. Also, the final S SKR period given by combining Ulysses/Cassini data displays significant long-term changes shifting from the simple smooth evolution of the planet inclination.


\begin{figure}[ht]
\centering
\includegraphics[width=\textwidth,angle=0]{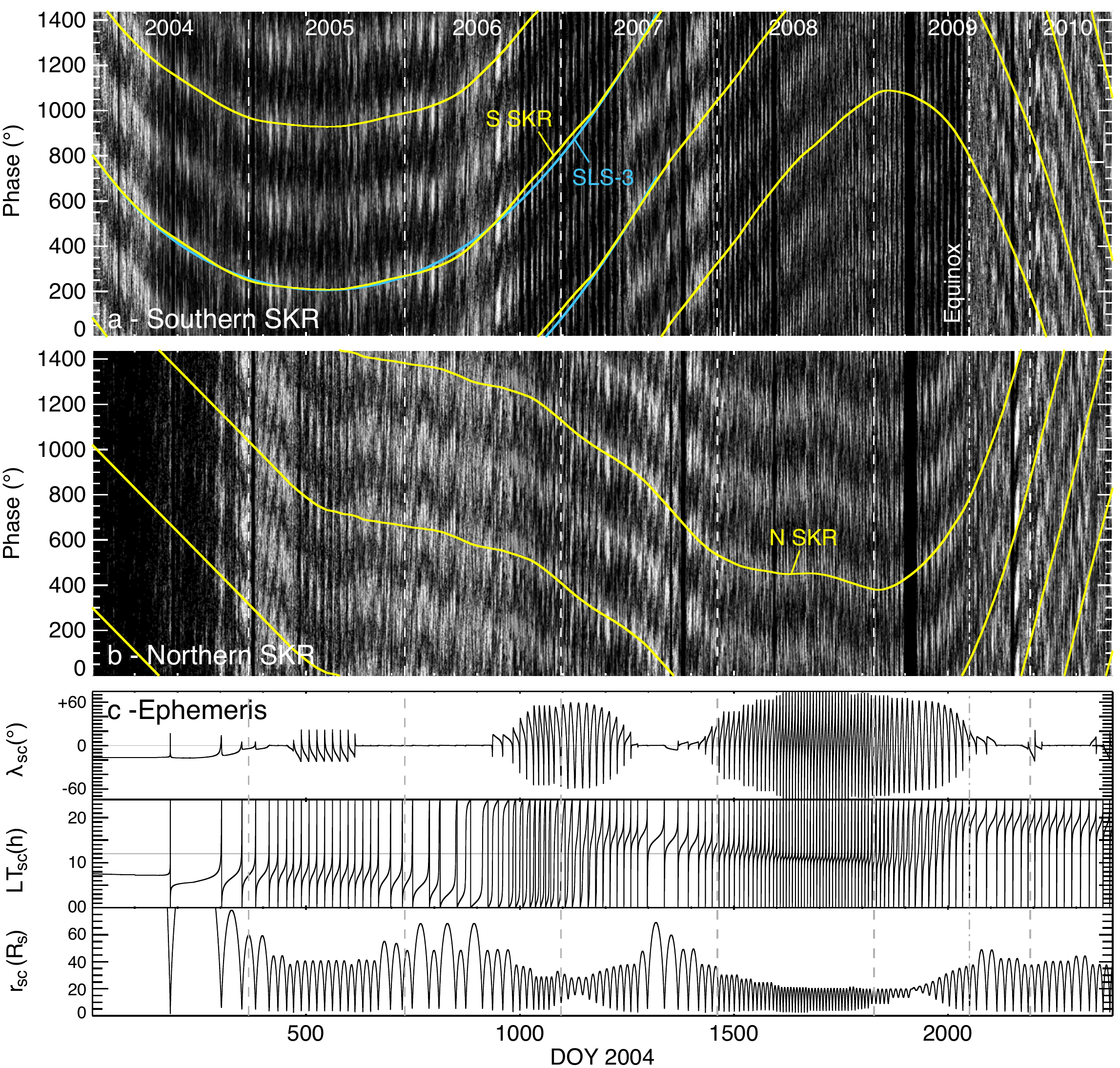} 
\caption{(a) Plot of $\log$~P$_\mathit{SKR,S}$(t) normalized to the average over one rotation as a function of an arbitrary phase computed from a fixed reference period of 10.7928~h (enabling direct comparisons with Kurth et al. [2008]), between 2004-001 and 2010-193. Each rotation is replicated four times along the y axis for clarity. The S SKR rotational modulation (given by $\Phi_\mathit{S}$(t)$~=0^\circ$) is displayed in solid yellow, while the SLS 3 one (given by sub-solar longitudes $=100^\circ$) is shown in solid light blue. Both correctly track S SKR maxima. (b) Identical plot for $\log$~P$_\mathit{SKR,N}$(t) and an arbitrary reference period of 10.6~h. (c) Cassini orbital parameters ($\lambda_\mathit{sc}$ the latitude, LT$_\mathit{sc}$ the local time and $r_\mathit{sc}$ the distance to the planet's center in kronian radii R$_\mathit{S}$, with 1~R$_\mathit{S}=60268$~km).}
\centering
\label{fig3}
\end{figure}

\subsection{Phase systems}
\label{phase}

Using continuous SKR periods, a separate phase system was built for each hemisphere from 1 January, 2004 (DOY 1) to 12 July, 2010 (DOY 2385). The phase of the SKR rotational modulation $\Phi_\mathit{S,N}$(t) was numerically integrated from:

\begin{equation}
\Phi_\mathit{S,N}(t)=\int{\frac{360}{T_\mathit{S,N}(t)}dt}+\Phi_\mathit{0,S,N}
\end{equation}

where t is the time and $\Phi_\mathit{0,S,N}$ an arbitrary reference. As $\Phi_\mathit{S,N}$(t) is defined modulo 360$^\circ$, SKR maxima occur at a fixed phase. $\Phi_\mathit{0,S,N}$ was chosen to reference SKR peaks at $\Phi_\mathit{S,N}(t)=0^\circ$, by fitting $\log$~P$_\mathit{SKR,S}(\Phi_\mathit{S})$ and $\log$~P$_\mathit{SKR,N}(\Phi_\mathit{N})$ (respectively averaged over the intervals 2004-2008 and 2005-2008, where rotational modulations are well defined) with cosine functions.

\begin{figure}[ht]
\centering
\includegraphics[width=\textwidth,angle=0]{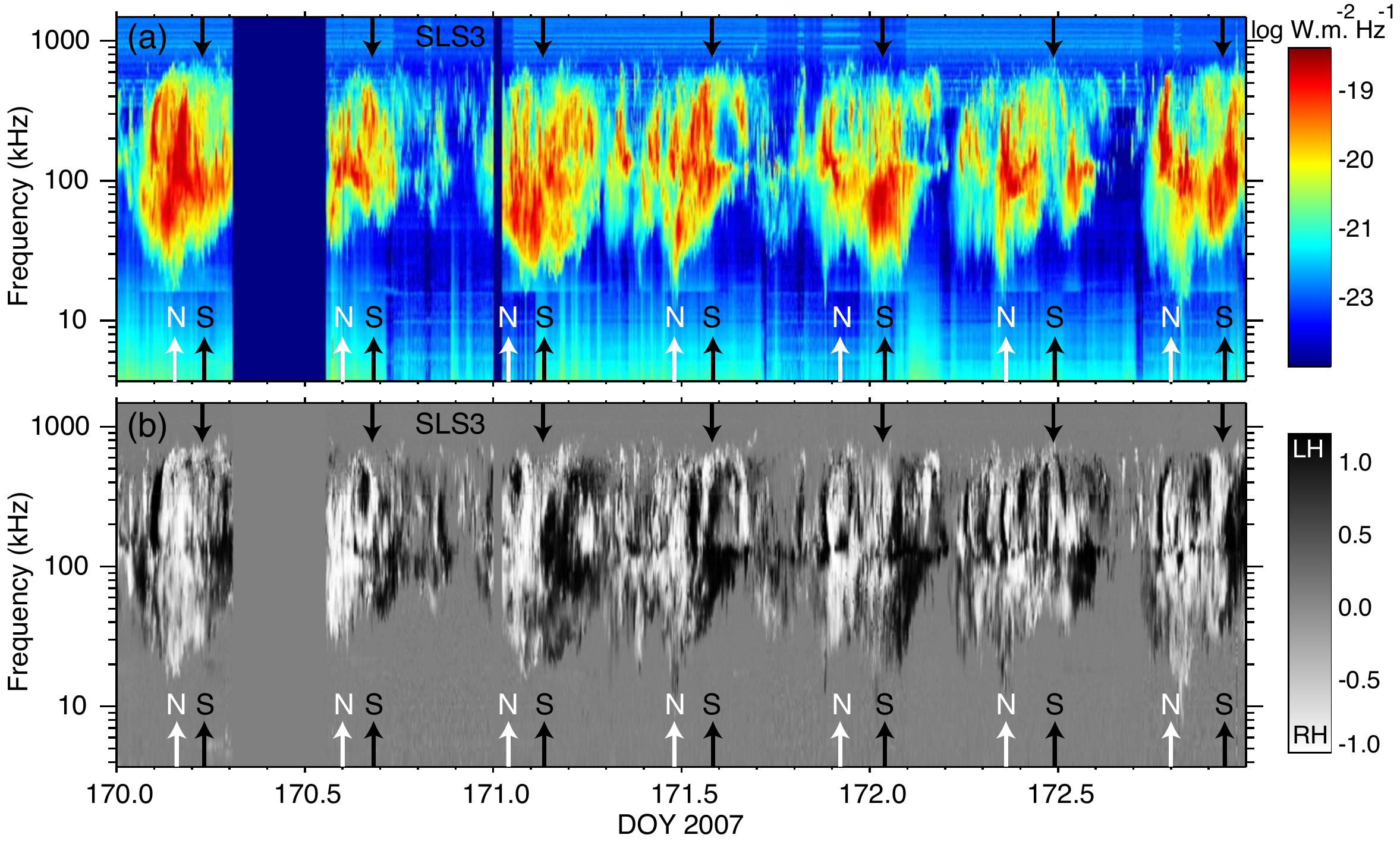} 
\caption{Dynamic spectra of (a) flux density (normalized to 1AU) and (b) degree of circular polarization for equatorial observations in 2007. LH (black) and RH (white) polarized bursts show the diurnal modulation of S and N X mode SKR. Top black arrows indicate SLS 3 SKR peaks. Bottom black and white arrows indicate S and N SKR peaks given by $\Phi_\mathit{S,N}$(t)~$=0^\circ$.}
\centering
\label{figsup}
\end{figure}

Figure \ref{fig3} illustrates the temporal evolution of S,N SKR phase drifts with respect to arbitrary constant rotation periods. Yellow solid lines show that the above defined SKR phases correctly track long-term SKR maxima for each hemisphere, with more noisy intervals after the equinox or at the end of 2007. However, the latter intervals correspond to near-equatorial observations from the dusk sector, from which the visibility of dawn radio sources can be questioned (see section \ref{modulation}), as suggested by previous modeling work [Lamy et al., 2008b] and direct observations [Cecconi et al., 2009]. $\Phi_\mathit{S}$(t) compares to the sub-solar longitude defined in previous SLS models. In Figure \ref{fig3}a, $\Phi_\mathit{S}$(t)$~=0^\circ$ and SLS 3 sub-solar longitude $=100^\circ$ (solid light blue) match on the interval where the latter is defined, and differ by generally less than 20$^\circ$ to 30$^\circ$. The relevance of the present phase systems is illustrated differently in Figure \ref{figsup} for 3 days of equatorial observations over which S and N SKR bursts (well identified in Figure \ref{figsup}b) match $\Phi_\mathit{S,N}$(t)$~=0^\circ$ (arrows).

The S and N phase systems have been used to organize successfully the power radiated by UV aurorae [Nichols et al., 2010] in both hemispheres, which demonstrated the existence of a SKR-like diurnal modulation of atmospheric aurorae, which had been unsuccessfully searched for a long time. In parallel, Andrews et al. [2010] identified two systems of magnetic field oscillations, interpreted as independent high-latitude field-aligned current systems, that pulse at S and N SKR periods to within $0.01\%$, with upward current layers matching SKR intense dawn sources during S and N SKR peaks. In addition, Provan et al. [2011] showed that part of the 'jitter' observed in equatorial magnetic oscillations can be explained by beating effects between both periods. 

\subsection{Mid-term variations}
\label{midterm}

To investigate fluctuations of SKR periods at shorter time scales, the above LS analysis was applied on the same dataset with a 100-days long sliding window. The resulting S,N periods are plotted in blue in Figure \ref{fig2}. Although less continuously defined, they display clear fluctuations of the order of a few months. Their correlation coefficient reaches $c=-0.41$, which increases to $c=+0.27$ once long-term trends (black) are subtracted. Considering the limited number of oscillations, their approximately one-to-one correspondence supports an overall correlation.

A symmetrical behaviour of $T_\mathit{S,N}$(t) may result from either orbital effects affecting the visibility of S and N SKR sources (whose intensity displays a clear LT dependence) or from a physical process affecting both hemispheres similarly. Orbital effects of the order of 15 to 30 days clearly appear in Figures \ref{fig3}a,b and prevent to track 20-30 days fluctuations previously identified by Zarka et al. [2007] in pre-SOI observations. However, no obvious orbital variations of the order of a few months are visible, which makes plausible a similar control of both SKR periods by a common physical cause. Checking the possible presence of similar fluctuations in other magnetospheric periodicities will help to determine their origin.


\section{Nature of the SKR diurnal modulation}
\label{modulation}

While most of periodic features observed in the kronian magnetosphere correspond to rotating phenomena, the SKR rotational modulation has been interpreted since the Voyager epoch as a strobe-like feature pulsing on the dawn sector. However, the radio source localization technique developed from the goniopolarimetric capability of Cassini radio instrumentation [Cecconi et al., 2009] has provided a new way to check this assumption. From this method, Lamy et al. [2009] have shown that SKR sources actually exist at all LT, forming a statistical radio oval conjugate to the UV one and whose intensity strongly varies as a function of LT, peaking at 08:00~LT. This result challenges the Voyager picture.

\begin{figure}[ht]
\centering
\includegraphics[width=\textwidth,angle=0]{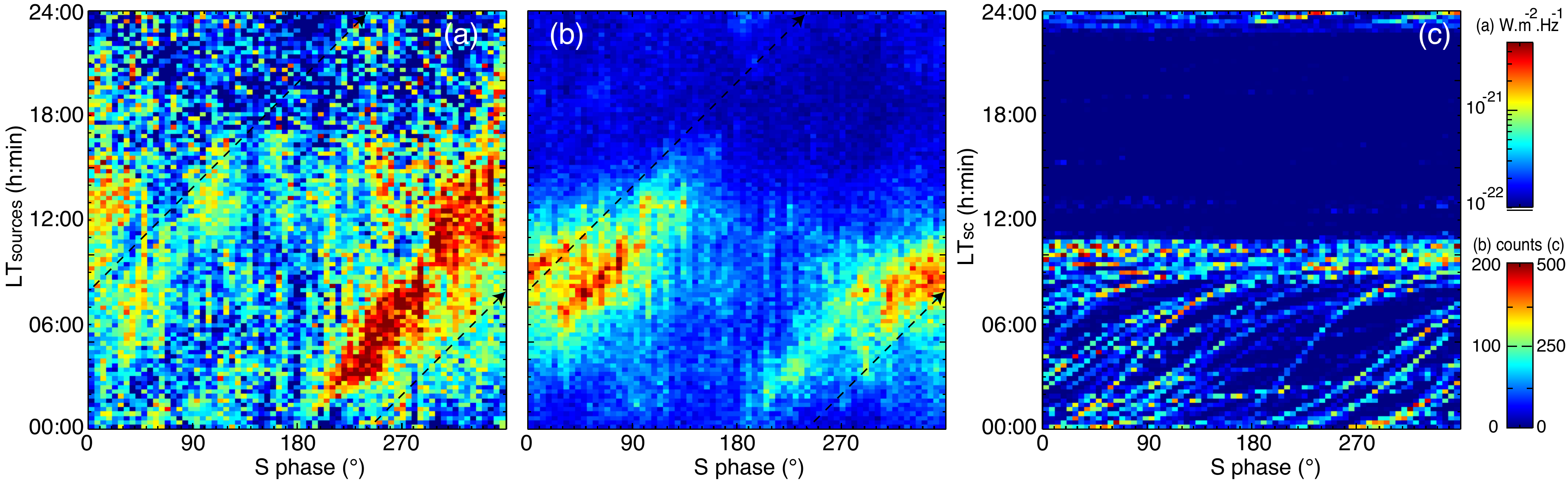} 
\caption{(a) Median flux density (normalized to 1AU) and (b) occurrence of S SKR sources organized as a function of their LT and $\Phi_\mathit{S}$ (in bins of 15~min by 5$^\circ$) from 2004-181 to 2010-193. The source LT was obtained from the radio source localization technique [Cecconi et al., 2009] applied to 3-antenna measurements for $r_\mathit{sc}\le20$R$_\mathit{S}$ and frequencies between 100 and 400~kHz. This dataset corresponds to 231 observing days, to which has been applied a standard data selection [Lamy et al., 2010] to sources with footprint latitudes between $-85^\circ$ and $-60^\circ$ (surrounding the S radio auroral oval). These selection criteria were satisfied by 1\% of the data, while the contribution of S sources detected after the equinox was negligible. The black dashed arrows materialize a source rotating at the S period, with a maximal intensity at 08:00~LT reached at $\Phi_\mathit{S}=0^\circ$. (c) Similar to (b) but as a function of the LT of the spacecraft and $\Phi_\mathit{S}$.}
\centering
\label{fig4}
\end{figure}

Here, $\Phi_\mathit{S}$(t) was used to organize the locus of S SKR sources (the most intense ones) derived from 6 years of individual 3-antenna measurements. They are less numerous than 2-antenna measurements, but yield unambiguous wave directions, and therefore accurate source locus, whatever the wave polarization (for details, see Fischer et al. [2009]). The spectral range was limited to the 100-400~kHz SKR spectral peak, for which intensities at different frequencies are both high and comparable. Figures \ref{fig4}a and \ref{fig4}b respectively show the median intensity and the occurrence of S SKR sources as a a function of their LT and $\Phi_\mathit{S}$. Figure \ref{fig4}c displays the associated orbital coverage with the occurrence of SKR sources organized as a function of the LT of the spacecraft and $\Phi_\mathit{S}$. Uncertainties on the radio sources position have not been taken into account in Figures \ref{fig4}a,b, that merely displays an average behaviour.

Figure \ref{fig4}c indicates that the S hemisphere was mainly visible when Cassini was between 23:00 and 11:00~LT, while Figure \ref{fig4}a illustrates that SKR sources were detected at all longitudes, with a clear LT dependence of their intensity, peaking around 08:00~LT. Then, whereas a clock-like pulsation in Figure \ref{fig4}a would correspond to highest intensities concentrated around $\Phi_\mathit{S}=0^\circ$, SKR sources surprisingly display a clear linear organization with LT($\Phi_\mathit{S}$), not seen in the orbital coverage (Figure \ref{fig4}c), and consistent with an intrinsically rotating phenomenon, materialized by the dashed arrows. But although the lowest intensities peak around $\Phi_\mathit{S}=180^\circ$, as expected, the highest ones are observed around $\Phi_\mathit{S}=300^\circ$, instead of $0^\circ$. This shift must be analyzed carefully as the intensity of detected sources merge contributions of a rotating feature, a strong LT dependence as well as important external effects (solar wind dynamic pressure, plasmoid activity) that can modulate the SKR flux by orders of magnitude. A complementary information is given by the occurrence of S SKR sources in Figure \ref{fig4}b, less sensitive to overall modulations, which peaks well around $\Phi_\mathit{S}=0^\circ$. The low statistics for dusk SKR sources is likely due to the poor dusk orbital coverage, together with the result of the point-source assumption used in the goniopolarimetric analysis (if a dawn bright source and a dusk faint source are observed together, the derived direction will correspond to the most intense - dawn - source). Finally, several linear features are present both left and right to the dashed arrows, suggesting that the longitudinal extent of the rotating pattern is significant and covers more than one quadrant in LT.

While this result needs to be separately confirmed by the symmetrical organization of SKR N sources (for which much less data fit the selection criteria before the equinox, both because of fainter intensity and poor visibility conditions) and the statistical analysis of 2-antenna measurements, it departs from the Voyager picture and reconciles the dynamics of SKR sources with the ones of other periodic magnetospheric phenomena. The above phase systems are reliable until the equinox in the sense that $\Phi_\mathit{S,N}$(t)$~=0^\circ$ indicate the transit of S,N rotating features in the dawn sector, about 08:00~LT. Indeed, the good organization of SKR maxima observed in Figure \ref{fig3}a,b before mid-2009 suggests that, despite severely changing visibility conditions, auroral radio sources of the active morning region have been detected by Cassini for most of the mission. 

\section{Conclusions}

In this article, accurate SKR periods were derived for each hemisphere from a 6-years long Cassini radio dataset analyzed by the Lomb-Scargle periodogram technique. S and N periods display clear anti-symmetrical yearly variations, consistent with a seasonal effect, together with symmetrical shorter term oscillations of the order a few months, whose origin remains to be identified. The continuous determination of S and N long-term varying periods, crossing 8 months after the equinox, was used to build phase systems for both hemispheres, valid from 1 January, 2004 (DOY 1) to 12 July, 2010 (DOY 2385), and freely available on the Cassini/RPWS/HFR data server: http://www.lesia.obspm.fr/kronos.

The S phase has been used to organize the locus of S SKR sources, which reveals an intrinsically rotating phenomenon. However, contrary to Jupiter, the search-light rotational modulation is affected by a strong LT dependence, which explains why the clock-like picture has remained a good approximation of the SKR periodic behaviour since Voyager.

\section*{References}
\everypar={\hangindent=1truecm \hangafter=1}


Andrews, D. J., A. J. Coates, S. W. H. Cowley, M. K. Dougherty, L. Lamy, 
G. Provan, and P. Zarka, Magnetospheric period oscillations at Saturn: 
Comparison of equatorial and high-latitude magnetic field periods with 
north and south SKR periods, \textsl{J. Geophys. Res.}, \textbf{115}, A12252, 2010.

Bonfond B., D. Grodent, J.-C. G\'erard, A. Radioti, J. Saur, and S. Jacobsen,
UV Io footprint leading spot: A key feature for understanding the UV Io 
footprint multiplicity?, \textsl{Geophys. Res. Lett.}, \textbf{35}, L05107, 2008.

Bunce, E. J., S. W. H. Cowley, D. L. Talboys, M. K. Dougherty, L. Lamy, 
W. S. Kurth, P. Schippers, B. Cecconi, P. Zarka, C. S. Arridge, and A. J. Coates,
Extraordinary field-aligned current signatures in Saturn s high-latitude 
magnetosphere: Analysis of Cassini data during Revolution 89, 
\textsl{Geophys. Res. Lett.}, \textbf{115}, A10238, 2010.

Carbary, J. F., D. G. Mitchell, S. M. Krimigis and N. Krupp, Dual 
periodicities in energetic electrons at Saturn, \textsl{Geophys. Res. 
Lett.}, \textbf{36}, L20103, 2009.

Cecconi, B. and P. Zarka, Model of a variable radio period for Saturn,
 \textsl{J. Geophys. Res.}, \textbf{110}, A12203, 2005.

Cecconi, B., L. Lamy, P. Zarka, R. Prang\'e, W. S. Kurth, and P. Louarn, 
Goniopolarimetric study of the rev. 29 perikrone using the Cassini Radio 
and Plasma Wave Science instrument high frequency radio receiver, 
 \textsl{J. Geophys. Res.}, \textbf{114}, A03215, 2009.

Desch, M. D., and M. L. Kaiser, Voyager measurement of the
rotation period of Saturn's magnetic field, \textsl{Geophys. Res. 
Lett.}, \textbf{8}, 253-256, 1981.

Fischer, G., B. Cecconi, L. Lamy, S.-Y. Ye, U. Taubenschuss, W. Macher, 
P. Zarka, W. S. Kurth, and D. A. Gurnett, Elliptical polarization of Saturn 
Kilometric Radiation observed from high latitudes, \textsl{J. Geophys. Res.}, 
\textbf{114}, A08216, 2009.

Galopeau, P. H. M., and A. Lecacheux, Variations in Saturn's radio
rotation period measured at kilometer wavelengths, \textsl{J. Geophys. Res.}, 
\textbf{105}, 13089--13101, 2000.


Gurnett, D. A. et al., Radio and Plasma Wave Observations at Saturn from Cassini's 
approach and first orbit, \textsl{Science}, \textbf{307}, 1255--1259, 2005.

Gurnett, D. A., A. M. Persoon, W. S. Kurth, J. B. Groene, T. F. Averkamp, 
M. K. Dougherty, D. J. Southwood, The Variable Rotation Period of the 
Inner Region of Saturn's Plasma Disk, \textsl{Science}, \textbf{316}, 442--444, 2007.

Gurnett, D. A., A. Lecacheux, W. S. Kurth, A. M. Persoon, J. B. Groene, 
L. Lamy, P. Zarka, and J. F. Carbary, Discovery of a north-south asymmetry in 
Saturn's radio rotation period, \textsl{Geophys. Res. Lett.}, \textbf{36}, L16102, 2009.

Gurnett, D. A., J. B. Groene, A. M. Persoon, J. D. Menietti, S.-Y. Ye, 
W. S. Kurth, R. J. MacDowall, and A. Lecacheux, The reversal of the rotational 
modulation rates of the north and south components of Saturn kilometric 
radiation near equinox, \textsl{Geophys. Res. Lett.}, \textbf{37}, L24101, 2010.

Horne, J. H. and S. L. Baliunas, A prescription for period analysis of unevenly sampled
time series, \textsl{Astrophys. J.}, \textbf{302}, 757-763, 1986.

Kaiser, M. L., M. D. Desch, J. W. Warwick, and J. B. Pearce,
Voyager detection of nonthermal radio emission from Saturn, \textsl{Science},
\textbf{209}, 1238-1240, 1980.

Kaiser, M. L., M. D. Desch, W. S. Kurth, A. Lecacheux, F. Genova, B. M. Pedersen, 
and D. R. Evans, Saturn as a radio source, in Saturn,
\textsl{Space Science Series}, edited by T. Gehrels and M. S. Matthews, 
Univ. Arizona Press, Tucson, Ariz., 378-415, 1984

Kurth, W. S., A. Lecacheux, T. F. Averkamp, J. B. Groene, and 
D. A. Gurnett, A Saturnian longitude system based on a variable 
kilometric radiation period, \textsl{Geophys. Res. Lett.},
\textbf{34}, L02201, 2007.

Kurth, W. S., T. F. Averkamp, D. A. Gurnett, J. B. Groene, and 
A. Lecacheux, An update to a Saturnian longitude system based 
on kilometric radio emissions, \textsl{J. Geophys. Res.},
\textbf{113}, A05222, 2008.

Lamy, L., P. Zarka, B. Cecconi, R. Prang\'e, W. S. Kurth, and 
D. A. Gurnett, Saturn kilometric radiation: Average and statistical 
properties, \textsl{J. Geophys. Res.}, \textbf{113}, A05222, 2008a.

Lamy, L., P. Zarka, B. Cecconi, S. Hess and R. Prang\'e, Modeling 
of Saturn kilometric radiation arcs and equatorial shadow zone, 
\textsl{J. Geophys. Res.}, \textbf{113}, A10213, 2008b.

Lamy, L., B. Cecconi, R. Prang\'e, P. Zarka, J. D. Nichols and 
J. T. Clarke, An auroral oval at the footprint of Saturn's kilometric 
radio sources, colocated with the UV aurorae, \textsl{J. Geophys. Res.},
\textbf{114}, A07201, 2009.

Lamy, L., P. Schippers, P. Zarka, B. Cecconi, C. S. Arridge, 
M. K. Dougherty, P. Louarn, N. Andr\'e, W. S. Kurth, R. L. Mutel,
D. A. Gurnett, and A. J. Coates, Properties of Saturn kilometric 
radiation measured within its source region, \textsl{Geophys. Res. Lett.},
\textbf{37}, L12104, 2010.

Lomb, N. R., Least-squares frequency analysis of unequally spaced data, 
\textsl{Astrophys. and Space Sci.}, \textbf{39}, 447-462, 1976.

Provan G., D. J. Andrews,	B. Cecconi, S. W. H. Cowley, M. K. Dougherty,	
L. Lamy and P. Zarka, Magnetospheric period magnetic field oscillations 
at Saturn: Equatorial phase jitter produced by superposition of southern-
 and northern-period oscillations, \textsl{J. Geophys. Res.}, in press.

Nichols, J. D. , B. Cecconi, J. T. Clarke, S. W. H. Cowley, J.~C. G\'erard, 
A. Grocott, D. Grodent, L. Lamy, and P. Zarka, Variation of Saturn's 
UV aurora with SKR phase, \textsl{Geophys. Res. Lett.}, \textbf{37}, 
L15102, 2010.

Scargle, J. D., Studies in astronomical time series analysis. II. Statistical
aspects of spectral analysis of unevenly spaced data, \textsl{
Astrophys. J.}, \textbf{263}, 835-853, 1982.

Seidelmann, P. K., V. K. Abalakin, M. Bursa, M. E. Davies, 
C. de Bergh, J. H. Lieske, J. Oberst, J. L. Simon, E. M. Standish 
and P. Stooke, et al., Report of the IAU/IAG working group on 
cartographic coordinates and rotational elements of the planets 
and satellites: 2000, \textsl{Celest. Mech. Dyn. Astron.}, \textbf{82}, 
No. 1, 83-111, 2002.

Ye S.-Y., D. A. Gurnett, J.B. Groene, Z. Wang and W. S. Kurth,
Dual periodicities in the rotational modulation of Saturn narrowband
emission, \textsl{J. Geophys. Res.}, \textbf{115}, A12258, 2010.

Zarka, P., L. Lamy, B. Cecconi, R. Prang\'e and H. O. Rucker,
Modulation of Saturn's radio clock by solar wind, \textsl{Nature}, 
\textbf{450}, 265-267, 2007.

\end{document}